\documentclass[reprint,preprintnumbers, amsmath,amssymb, aps]{revtex4-2}

\usepackage{graphicx}
\usepackage{dcolumn}
\usepackage{bm}
\usepackage{changes}
\usepackage{color}

\begin{document}

\title{Unidirectional polarization beam splitters via exceptional points and finite periodicity of Non-Hermitian PT-symmetry}
\author{Jeng Yi Lee}
\affiliation{Department of Opto-Electronic Engineering, National Dong Hwa University, Hualien 974301, Taiwan}

\date{\today}

\begin{abstract} 
We present a theoretical study of a  novel polarization beam splitter (PBS), different to conventional time-reversal symmetry one,  where can be totally reflected at two opposite sides with one specific linearly polarized light incident and can be transparent at only one side with its orthogonal linearly polarized light incident.
The mechanism we employ is by both an exceptional point and finite periodicity of Non-Hermitian PT-symmetry.
To be more specific, we design such PBS made of a finite periodic structure in which each unit cell has a delicate balance gain and loss in spatial distribution. 
In order to have one linearly polarized light totally reflected, the corresponding polarized unit cell has to be operated at some PT-symmetry phase of reflection band.
To have single-sided transparent for its orthogonal polarized light, the corresponding polarized unit cell should be designed at an exceptional point as well as has asymmetric reflectance.
Interestingly, such single-sided transparent phenomenon is independent of total number of unit cell.
We believe this asymmetry PBS  may excite a new route to polarization control.

\end{abstract}
\pacs{ }

\maketitle

\section{Introduction}

Polarization control is a vital in numerous applications including optical communication, distinguishing chirality of molecules,  polarization holography to name a few \cite{p1,p2,p3,p4,p5,p6,p7,p8,p9,p10,p12}.
For example, introducing a degree of freedom by two orthogonal polarized light to transmitted signals, both capacity and spectral efficiency (SE) per unit bandwidth can be enhanced. 
In addition, many   molecules have structural chirality,  differently responding left- and right- circular polarized light.
Holography utilized by polarization of light can increase capacity of information record and data storage, in which can be achieved by a spatial array of ellipse-shaped silicon with different  orientation \cite{p12}.

An effective polarization control can be resolved by polarization beam splitter (PBS), that can decompose any incoming beam into
two/multiple orthogonal polarized beams.
Among many PBS architectures for various demands,  one type has devoted toward miniaturization, integration with other devices, and planar configuration, that is achieved by subwavelengthly phase-shifting arrays, i.e., metasurfaces \cite{alu1,linear1,linear2,linear3,linear4,linear5,p11,p12}.
However, the solution to dynamic polarization control is of more interest and challenge.
One approach for a linearly polarization rotator can be achieved by a chiral metasurface operated at a coherent perfect absorption-like condition where is incident with two counter-propagating linear poliarized light with tuning relative phases \cite{p11,dynamics1,dynamics2}.

Polarization beam splitters (PBS) assisted by photonic crystal heterostructure can split any incoming beams to two orthogonal polarized beams \cite{photonic_crystal1,photonic_crystal2},  by an operation of one linearly polarized beam at forbidden band and its orthogonal polarized beam at transmission band. 
However, time-reversal symmetry poses a fundamental constraint to these devices, resulting in symmetric reflectances at two illuminating sides and the corresponding scattering matrix being unitarity.


On the other hand, an unique asymmetric reflection based on Non-Hermitian parity-time  (PT) symmetry, 
offers a new architecture for this polarization control \cite{PT1,PT2,PT3}.
When arbitrary PT-symmetry systems are manipulated at exceptional point (EP), the scattering of systems can exhibit  transparent at one side while at another side  it would experience partial reflection as well as transmission.
However, in order to split any incoming beam into two orthogonal linearly polarized beams as well as guide two splitted beams to separated paths, such EP condition is not sufficient.
To address this issue, we employ a finite periodic structure  where each unit cell is made of a balance gain and loss.
With one specific linearly polarized beam incident the system is at reflection band, while with its orthogonal polarized beam incident that is at EP of transmission band and has asymmetry reflection.
At reflection band, two-sided reflectances are formed a pair of reciprocal moduli.
Interestingly, such one-sided transparent phenomenon could be established for any number of unit cell, as long as the unit cell is at EP as well as has asymmetric reflection.
We numerically verify such PBS via COMSOL Muliphysics. 
The proposed PBS via extraordinary non-Hermitian PT-symmetry may become a new component for relevant polarizing photonic systems.

\begin{figure*}[ht]
\centering
\includegraphics[width=0.9\textwidth]{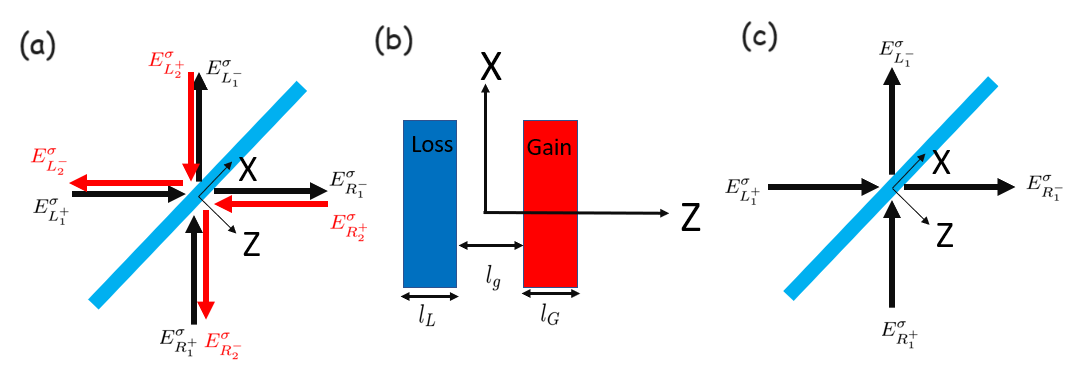}
\caption{(a) Schematic of a polarization beam splitter, constructed by a finite periodic structure with N-unit cells embedded. Each unit cell is made of gain-gap-loss, as shown in (b). The refractive indices for gain and loss are $n_G$ and $n_L$, respectively. The geometric thicknesses for gain, loss, and gap slabs are $l_G$, $l_L$, and $l_g$, respectively. At each polarization excitation, there are four independent incoming waves to the system, i.e., $\{E^{\sigma}_{L_1^{+}}, E^{\sigma}_{L_2^{+}},E^{\sigma}_{R_1^{+}},E^{\sigma}_{R_2^{+}}\}$, with relevant four outgoing waves, i.e., $\{E^{\sigma}_{L_1^{-}}, E^{\sigma}_{L_2^{-}},E^{\sigma}_{R_1^{-}},E^{\sigma}_{R_2^{-}}\}$. Here $\sigma$ denotes s-(TE) or p-(TM) polarization, $L$ and $R$ mean left and right sides, $+$ and $-$ mean incoming and outgoing waves, and $1$ and $2$ denote two identical scattering events with same incident angles. Our polarization beam splitter  is in general a four-port scattering system at each polarization excitation. We note that the wave propagating vectors are marked by red and black arrows. To simulate most situations, we consider only one incoming wave at each side, in association with related outgoing waves as shown in (c). } 
\end{figure*}

\section{Theory}
Figure 1 (a) depicts our PBS system, constructed by finite periodic PT-symmetry slabs.
The unit cell is made of homogeneous and isotropic media with gain, gap, and loss, illustrated in Fig. 1 (b). 
It satisfies PT-symmetry, that the refractive indices of lossy being $n_L$ and gain being $n_G$ meet $n_L=n^{*}_G$ and the thicknesses for lossy being $l_L$ and for gain being  $l_G$ are same, i.e., $l_L=l_G=l_s$.
The thickness of gap is dented as $l_{g}$.
Throughout this work, we let the refractive index of environment be $n=1$.
We discuss that incident waves obliquely illuminate the PBS.
In this situation, our PBS system is in general a four-port wave scattering  allowing four independent incoming waves and related outing waves at each s- (TE-) and p- (TM-) polarization excitation.
Here s- polarization and p-polarization are denoted as oscillation of electric fields and magnetic fields being orthogonal to a plane of incidence, respectively. 
Moreover, the complex amplitudes for incoming waves in our system are $\{E^{\sigma}_{L_1^{+}}, E^{\sigma}_{L_2^{+}},E^{\sigma}_{R_1^{+}},E^{\sigma}_{R_2^{+}}\}$ and the complex amplitudes for the corresponding outgoing waves are  $\{E^{\sigma}_{L_1^{-}}, E^{\sigma}_{L_2^{-}},E^{\sigma}_{R_1^{-}},E^{\sigma}_{R_2^{-}}\}$.
Here the notations of $+$ and $-$ are denoted as  incoming and outgoing waves, the subscripts of $L$ and $R$ are denoted as the left and right sides, the superscript of $\sigma$ is denoted as s- or p- polarization, and $1$ and $2$ are marked for two identical scattering situations with same incident angles.
Toward most situations, we consider one incoming wave at each side, as shown in Fig. 1 (c).
The scattering matrix related to incoming and outgoing waves at each polarization excitation can be expressed by
$\overline{S}^{\sigma}\vert +^{\sigma}>=\vert -^{\sigma}>$, where $\vert +^{\sigma}>$ is   polarization-dependent incoming wave vector expressed as $\vert +^{\sigma}>=[E^{\sigma}_{L_1^{+}},E^{\sigma}_{R_1^{+}}]^T$ and $\vert -^{\sigma}>$ is  polarization-dependent outgoing wave vector expressed as $\vert -^{\sigma}>=[E^{\sigma}_{L_1^{-}},E^{\sigma}_{R_1^{-}}]^T$. 
This scattering matrix of $\overline{S}^{\sigma}$ is polarization dependent.
The discussion of scattering matrix under the oblique illumination is placed at Appendix A.

\begin{figure*}[ht]
\centering
\includegraphics[width=1\textwidth]{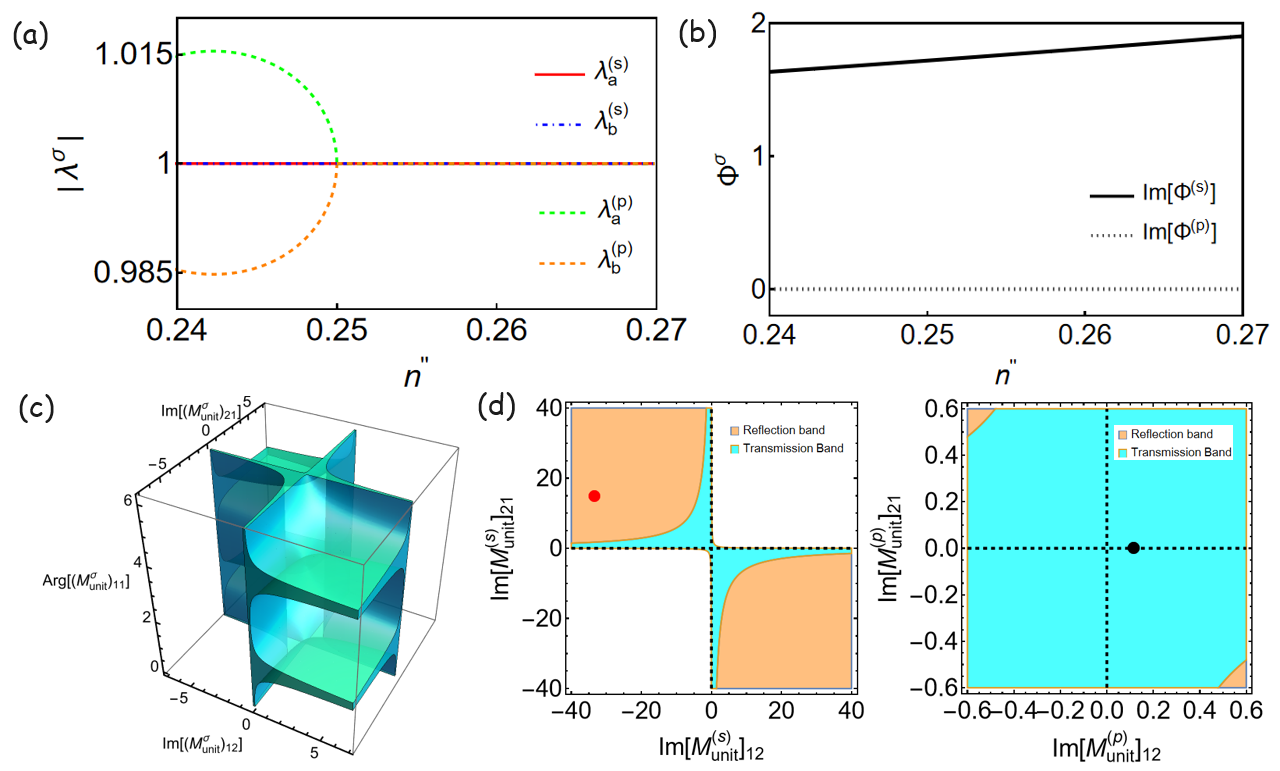}
\caption{We calculate the scattering eigenvalues of our designed unit cell with respect to $n^{''}$ as shown in (a).
There are two eigenvalues at each polarizing excitation, denoted by the subscripts of a and b.
We also calculate the Bloch phase with respect to $n^{''}$ in (b), which indicates that the s-polarizing unit cell is at reflection band while the p-polarizing cell is at transmission band. 
With a consideration of generic PT-symmetry constraint and reciprocity principle, there are only three degrees of freedom to the transfer matrix of arbitrary PT-symmetry unit cells at each polarization excitation. 
Then, with further consideration of Bloch phase, we depict a generic 3D parametric space to display allowable independent parameters, enabling the formation of wave transmission band in (c).
By taking $Arg[(M^{(s)}_{unit})_{11}]=4.84$ and $Arg[(M^{(p)}_{unit})_{11}]=5.79$, we plot 2D parametric spaces in terms of  $Im[(M^{\sigma}_{unit})_{12}]$ and $Im[(M^{\sigma}_{unit})_{21}]$ for each polarization excitation, where the regions supporting transmission band and reflection band are marked by cyan and orange colors, respectively. 
The corresponding scattering of our unit cell is marked by a red dot and a black dot in (d).
 In the left panel, we find the s-polarizing unit cell at PT symmetry phase as well as reflection band. 
 In the right, we find the p-polarizing unit cell at EP as well as transmission band. The black dashed lines correspond to EP. } 
\end{figure*}

The scattering properties of any PT-symmetry photonics can be categorized into PT symmetry phase, exceptional point (EP), and PT broken symmetry phase, which can be simply judged by transmittances with $T<1$, $T=1$, and $T>1$, respectively \cite{yidong1,yidong2,our1,our2}.
More interestingly, using these PT-symmetric systems as a building block (unit cell), the finite periodic systems would form wave transmission band or reflection band as well as encounter PT phase transition \cite{finite1,our2}.

The condition to determine the formation of transmission band or reflection band is by
\begin{equation}\label{pt3}
\cos\Phi^{\sigma}=\text{Re}(1/t^{\sigma}_1),
\end{equation}
where $t^{\sigma}_1$ is complex transmission coefficient of unit cell and  $\Phi^{\sigma}$ corresponds to the Bloch phase in infinite periodicity.
 Eq.(\ref{pt3}) is polarization dependent, denoted by a superscript of $\sigma$.
As $\Phi^{\sigma}$ becomes a complex value or pure real value, it corresponds to the formation of reflection band or transmission band, respectively.
We point out that within a framework of PT-symmetry, the transmittance at transmission band may be larger than unity.
At reflection band, the transmittance is indeed zero, while the reflectances may be higher or lower than unity.
Combining the transmission property of a unit cell with  Eq.(\ref{pt3}), we can deduce that arbitrary unit cells having EP and PT broken symmetry phase would always lead to have transmission band for any N, while that having PT symmetry phase would lead to have transmission band or reflection band, depending on the transmission phase of unit cell \cite{finite1,our2}.

On the other hand, the transmittance between a finite periodic system and its unit cell can be resolved by
\begin{equation}\label{pt4}
1-\frac{1}{T^{\sigma}_N}=(1-\frac{1}{T^{\sigma}_1})\frac{\sin^2[N\Phi^{\sigma}]}{\sin^2\Phi^{\sigma}},
\end{equation}
derived from the work of \cite{finite1}.
Again, this is polarization-dependent expression.
Here $N$ is total number of unit cells, $T^{\sigma}_1=\vert t^{\sigma}_1\vert^2$ is polarization-dependent transmittance of unit cell, and $T^{\sigma}_N$ is  polarization-dependent transmittance for N-cells.
At transmission band, the Bloch phase is real.
Then we analytically find that as any unit cells having PT symmetry phase, it would result in $T^{\sigma}_N\leq 1 $, corresponding to perform PT symmetry phase and EP for any N.
We mark that such EP result occurred at N-cells in absence of the unit cell at EP would accompany with bi-directional transparency at both sides \cite{our2}.
Moreover, as any unit cells having EP, it would always have EP for any N cells, i.e., $T^{\sigma}_N=1$.
This case, if the unit cell also has asymmetry reflection, it would lead to have asymmetry reflection occurred at N-cells.
When any unit cell is at PT broken symmetry phase, it would have PT broken symmetry phase or EP for any N, owing to $T^{\sigma}_N\geq 1$.
Again, such EP result would have bi-directional transparency at both sides.
Therefore, the approach to support an unidirectional transparency for our desirable PBS is only when the unit cell is operated at EP and has asymmetric reflection. 

\begin{figure*}[ht]
\centering
\includegraphics[width=1\textwidth]{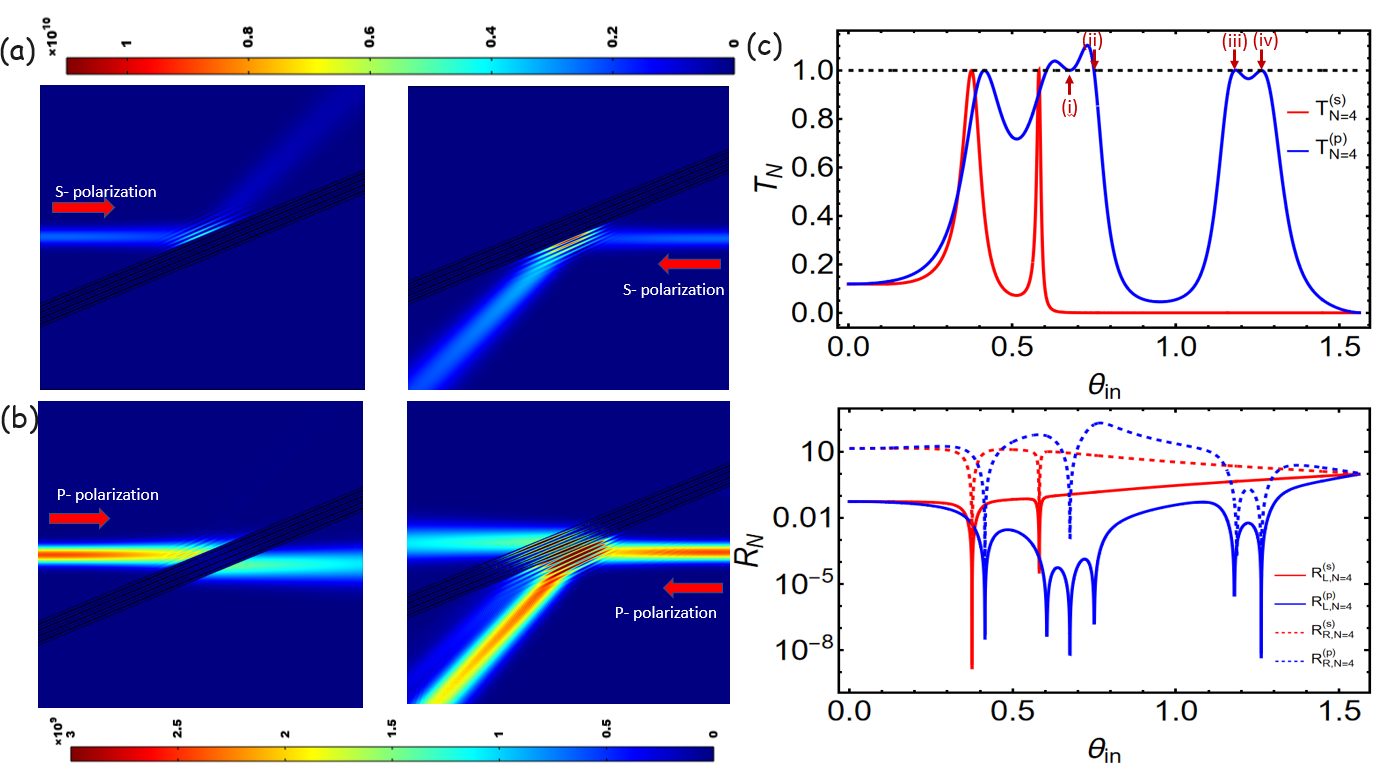}
\caption{ A Gaussian beam is used to obliquely illuminate our PBS at two sides at each polarization, in (a)-(b). Here we show the corresponding absolute values of electric fields. The parameters of this PBS are listed in the first column of Table 1. We can observe total reflection occurred at s-polarizing excitation, also with asymmetry reflection.  At p-polarizing excitation, the importing Gaussian beam becomes transparency at the left side, but it would encounter reflection at the right side.
In order to understand the scattering  of this single-sided PBS with respect to a variety of incident angles $\theta_{in}$, we thus calculate the corresponding transmittance and reflectances in (c). In the upper panel, the unity transmittances at p-polarization excitation are found at four incident angles, marked by (i)-(iv), corresponding to the  EP emerged.
Meanwhile, at these incident angles, the s-polarizing unit cells are  operated at reflection band. 
In the bottom panel, we can observe asymmetry reflection at the cases of (ii), (iii), and (iv).
In the case of (i), it has zero reflections at two sides.} 
\end{figure*}

\begin{table*}[ht]
\caption{\label{tab:table1} Parameters of unidirectional PBS}
\begin{ruledtabular}
\begin{tabular}{cccc}
& Systematic configuration \# 1  &  Systematic configuration \# 2  &  Systematic configuration \# 3  \\ \hline
Incident angle, $\theta_{in}$ & $1.179$ & $1.088$ & $0.796$\\
Operating wavelength & $620[nm]$ & $600[nm]$ & $580[nm]$ \\
 $l_s$ & $ 0.8 \lambda$& $0.8 \lambda$& $0.7 \lambda$\\
$l_g$&$118.26 [nm]$& $257.658 [nm]$& $133.89[nm]$\\
$n_L$ &$2 +  0.25i$ & $1.85 +  0.2i$ & $2.2+0.15i$\\
Number of unit cell, $N$ &4&6&6\\
 $T_{N}^{(s)}$   & $6.41714\times 10^{-8}$& $9.854\times 10^{-9}$ &$5.36485\times 10^{-5}$\\
 $T_{N}^{(p)}$   &1& 1&1\\
 $R^{(s)}_{L,N}$   &0.444138&0.403082& 0.152341\\
 $R^{(p)}_{L,N}$   & $4.41097\times 10^{-30}$& $2.08179\times 10^{-26}$& $4.52699\times 10^{-28}$\\
$R^{(s)}_{R,N}$  &2.25155&2.48089& 6.56352\\
 $R^{(p)}_{R,N}$  &0.0518471& $3.03\times 10^{-4}$ & 84.7401\\
\end{tabular}
\end{ruledtabular}
\end{table*}

\section{Design of polarizing beam splitters}
Based on above analysis, we begin to design the unit cell.
Firstly, we use these parameters by operating wavelength being $620 [nm]$, incident angle being $\theta_{in}=1.179$, thickness of slab being $l_s=0.8\lambda$, and thickness of gap slab being $l_g=118.26 [nm]$, and $n_L=2+in^{''}$. 
Here $n^{''}$ is a tunable parameter related to material loss.
Then, by scanning $0.24\leq n^{''}\leq 0.27$, we calculate the corresponding scattering eigenvalues at s- and p- polarization excitation in Fig. 2 (a), to distinguish PT phase.
At each polarization excitation, there are two scattering eigenvalues that are $\lambda^{\sigma}_a=t^{\sigma}_1+\sqrt{r^{\sigma}_{L,1}r^{\sigma}_{R,1}}$ and $\lambda^{\sigma}_b=t_1^{\sigma}-\sqrt{r^{\sigma}_{L,1}r^{\sigma}_{R,1}}$. Here
$r^{\sigma}_{L,1}$ and $r^{\sigma}_{R,1}$ represent polarization-dependent reflection coefficients of a unit cell by left and right incidences.
In s-polarization excitation,  we numerically find the s- polarizing unit cell at PT symmetry phase, as in Fig. 2 (a).
However,  in p-polarization excitation, we can observe that the unit cell is at PT broken symmetry phase by $0.24\leq n^{''}<0.25$, at PT symmetry phase by $0.25<n^{''}\leq 0.27$, and at EP by $n^{''}=0.25$.
We further calculate the Bloch phase of Eq.(\ref{pt3}) with respect to $0.24 \leq n^{''}\leq 0.27$ as shown in Fig. 2 (b).
The calculation indicates that  the p-polarizing unit cell is at transmission band, while the s-polarizing unit cell is at reflection band.

In order to understand an integrated information between Bloch phase and PT phase of unit cells, we draw a support by parametrization applied to the transfer matrix of unit cell \cite{our1,our2}.
Due to a PT-symmetry constraint to the transfer matrix, the degree of freedom is only three \cite{long1}.
We thus adopt $Im[(M_{unit})_{12}]$, $Im[(M_{unit})_{21}]$, and $Arg[(M_{unit})_{11}]$ as independent real parameters. 
Moreover, under reciprocity principle, there has another constraint by $Im[(M_{unit})_{12}]Im[(M_{unit})_{21}]\leq 1$ \cite{our1,long1}.
In the meanwhile, with a consideration of Bloch phase of Eq.(\ref{pt3}), we depict the generic 3D parametric space to exhibit wave transmission for any periodic PT-symmetry photonics, beyond any system configurations, operating frequencies, and materials, as green-blue colors of Fig. 2 (c).

By taking $n^{''}=0.25$, we numerically obtain the polarization-dependent transmission phases for the unit cell with $Arg[(M^{(s)}_{unit})_{11}]=4.84$ and $Arg[(M^{(p)}_{unit})_{11}]=5.79$.
Consequently, we depict 2D parametric spaces in terms of $Im[(M^{\sigma}_{unit})_{12}]$ and $Im[(M^{\sigma}_{unit})_{21}]$ to show more detailed information with PT phases, transmission band, and reflection band in Fig. 2 (d). 
The region marked by cyan color corresponds to transmission band. 
The region with orange color means reflection band.
Moreover, the region supporting PT symmetry phase is located at the colored second and fourth quadrants excluding white region and black dashed lines, while the region supporting PT broken symmetry phase is located at the colored first and third quadrants also excluding white region and black dashed lines.
In between PT symmetry and broken symmetry phases, it corresponds to EP, marked by black dashed lines.
The corresponding scattering of our  unit cell is marked by red and black dots in Fig. 2 (d).
Obviously, the p-polarizing unit cell is at EP as well as transmission band, while the s-polarizing unit cell is at PT symmetry phase as well as reflection band.

 We then employ such unit cell to build our PBS and choose $N=4$ cells.
Now with COMOSL Multiphysics, a Gaussian beam with linear polarization is used to illuminate the PBS from each side as in Figs.3 (a)-(b).
We can see that at s-polarization excitation, the Gaussian beams are totally reflected at both sides, also with asymmetry reflection.
The strength of left reflected beam smaller than right one is due to that the gain and the loss slabs are placed at the left and the right in our case.
At p-polarization excitation, we can see unidirectional transparent at the left, while it encounters reflection at the right.
Next, our interest is to understand the scattering of our PBS with respect to any incident angles.
We then calculate the total transmittance of $T_{N}^{\sigma}$ and two total reflectances of $R_{L,N}^{\sigma}$ and $R_{R,N}^{\sigma}$ in Fig. 3 (c). 
We observe that there are four incident angles with $\theta_{in}=[0.675,0.75,1.179,1.186]$ marked by (i)-(iv) with the unity transmittance at p-polarization excitation.
In reflectance, we observe that the cases of (ii)-(iv) would have one-sided zero reflection, while the case of (i) has double-sided zero reflection.
To understand the PT phases of the unit cell at these incident angles, we utilize the 2D parametric space as a same approach in Fig. 2 (d).
The results are placed at Appendix B.
 We find that the p-polarizing unit cell in the cases of (ii)-(iv) would be at EP and have asymmetry reflection.
However, in the case of (i), the p-polarizing unit cell is not at EP, instead at PT broken symmetry phase.
Such case satisfies another EP condition by $\sin[N\Phi^{\sigma}]=0$, resulting in double-sided transparency.
We note that this double-sided transparency phenomenon operated at such incident angle is N-dependent.

With advances in nano-scale fabrication and development of material sciences, nano-structured gain materials operated at visible spectra can be achieved by GaAs quantum wells,  MAPbI3 perovskite, CdS quantum dots, rare earth ions doped materials (Eu3+, Sm3+, Dy3+), and  fluorescent polymer \cite{active1,active2,active3,active4,active5}.
These materials provide an opportunity to realize our proposed PBS. 
Accordingly, we provide another two sets of system configuration in table I that exhibit the realizable architecture for our proposed PBS at possible operating conditions.

 \begin{figure}[ht]
\centering
\includegraphics[width=0.5\textwidth]{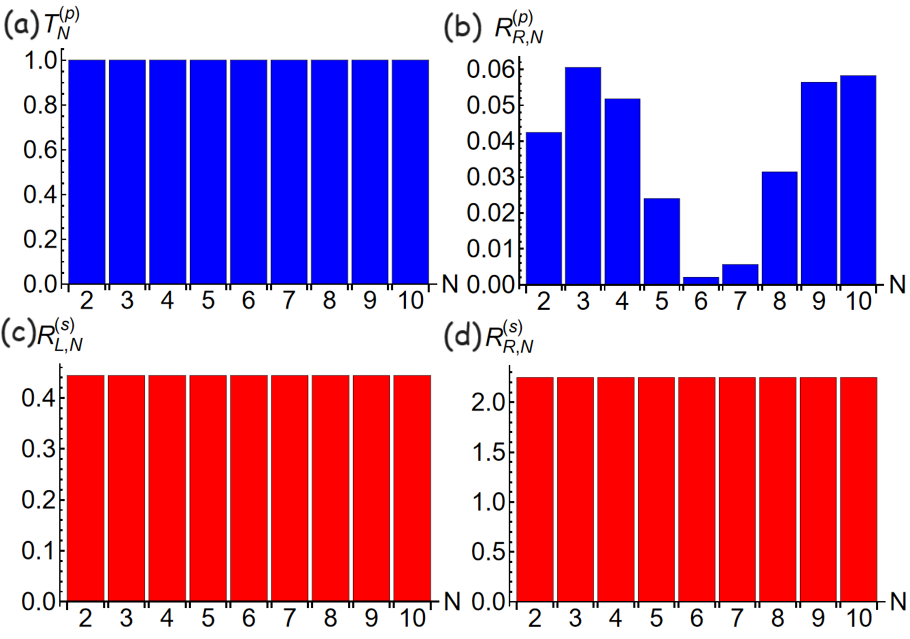}
\caption{ Calculation of transmittance and reflectances with respect to N at s- and p- polarizing excitations. In (a), as the p-polarizing unit cell is at EP and has single-side transparency, the finite periodic system would exhibit N-independent unity transmittance. However, the corresponding reflectance at right side would have N-dependence in (b). At reflection band,  the  reflectances at two sides have a pair with reciprocal moduli as shown in (c)-(d). Here the parameters used in the unit cell are listed in the first column of Table 1. } 
\end{figure}

\section{Generic Reflectance relation}
We discuss the reflectance relation for our proposed PBS.

In our case, the emgerency of EP occurred at the p-polarizing unit cell leads to have $T_1^{(p)}=1$. 
Accordingly, using Eq. (\ref{pt4}), we find $T_N^{(p)}=1$, independent of N.
This result can get a support by numerical calculation of Fig. 4 (a). 
Next, based on the outcome of Ref. \cite{finite1}, we can formulate the reflectance relation between a finite periodic system and its unit cell,   
\begin{equation}\label{pt5}
R_{R,N}^{(p)}=R_{R,1}^{(p)}(\frac{\sin[N\Phi^{(p)}]}{\sin[\Phi^{(p)}]})^2.
\end{equation}
Obviously, the total reflectance of finite periodic system depends on N.
In Fig. 4 (b), the numerical results support this finding.

On the other hand, at s-polarization excitation, due to the operation of reflection band, i.e., $T^{(s)}_{N\rightarrow \infty}\rightarrow 0$, we obtain reflectance and transmittance relations between a  finite periodic system and its unit cell
\begin{equation}
\begin{cases}
\begin{split}
&R^{(s)}_{R,N}R^{(s)}_{L,N}=1\\
&R^{(s)}_{R,N}=\frac{R^{(s)}_{R,1}}{1-T^{(s)}_1}.
\end{split}
\end{cases}
\end{equation}
The derivation is placed at Appendix C.
Here we can see that the total reflectances at two sides are N-independent.
Moreover, their reflectances form a pair of reciprocal moduli, as numerically supported in Figs. 4 (c)-(d).

Last but no least, we should remark that the single-sided PBS is owing to that the unit cell is at EP as well as has asymmetry reflection.
Such unique uni-directional transparent is N-independent, as numerically supported in Fig. 4 (a).
The role of N would affect the reflection at non-transparent side, as in Fig. 4 (b).
In addition, we note that there could have "accidental" bi-directional transparency occurred at some  unit cells.
However, using such unit cell as a building block, the PBS would become bi-directional transparent \cite{yidong2}.
On the other hand, another solution to support bi-directional transparent is by N-cells satisfying  $\sin[N\Phi^{\sigma}]=0$ \cite{our2,reflectionless1,reflectionless2}, already observed in the case of (i) of Fig. 3 (c).

Utilizing other systematic configurations, materials, operating frequencies, it is possible to have p-polarizing unit cells at reflection band and s-polarizing ones at EP, different to our architecture.
In acoustic, PT-symmetry beam splitters having single-sided transparent have implemented by proper impedance modulation of waveguide sidewall \cite{acoustic1,acoustic2}.

\section{Conclusion}
To our summary,  we propose a novel beam splitter with total reflection by specific linearly polarized beam incident, and with uni-directional transparency by its orthogonal polarized beam incident.
The mechanism is based on EP and finite periodicity of non-Hermitian PT symmetry.
In our architecture, we design the polarization-dependent unit cell for transparent beam at EP and having asymmetry reflection, while that for totally reflected beam at PT-symmetry phase of reflection band.
We also discuss the reflectance relations at transmission and reflection bands.
At reflection band, they form a  pair of reciprocal moduli, different to conventional time reversal symmetry PBS.
Our novel uni-directional PBS may provide exciting applications in relevant polarizing photonic systems.

\section*{Acknowledgements}

This work was supported by Ministry of Science and Technology, Taiwan (MOST) ($112$-$2112$-M-$259$ -$001$ -).

\section*{Appendix A}
At S-polarization (TE polarization) of plane wave excitations, the total electric waves at the left and right sides can be written as
\begin{equation}
\begin{split}
\vec{E}^{(s)}_L &=\hat{y}(E^{(s)}_{L_1^{+}} e^{i(\vec{k}_{L,in}-\omega t)}+E^{(s)}_{L_1^{-}}  e^{i(\vec{k}_{L,r}-\omega t)})\\
\vec{E}^{(s)}_R &=\hat{y}(E^{(s)}_{R_1^{-}}  e^{i(\vec{k}_{R,r}-\omega t)}+E^{(s)}_{R_1^{+}}  e^{i(\vec{k}_{R,in}-\omega t)})
\end{split}
\end{equation}
where $\vec{k}_{L,in}$ and $\vec{k}_{R,in}$ are denoted as incident wave vector at the left and right sides, $\vec{k}_{L,r}$ and $\vec{k}_{R,r}$ are denoted as reflected  wave vector at the left and right sides.
We note that due to same environments at the right and left sides, we have $\vec{k}_{L,i}=\vec{k}_{R,r}$ and $\vec{k}_{L,r}=\vec{k}_{R,i}$.

At P-polarization (TM polarization) excitation, the total magnetic waves at the left and right sides can be expressed by
\begin{equation}
\begin{split}
\vec{H}^{(p)}_L &=\hat{y}(H^{(s)}_{L_1^{+}} e^{i(\vec{k}_{L,i}-\omega t)}+H^{(s)}_{L_1^{-}}  e^{i(\vec{k}_{L,r}-\omega t)})\\
\vec{H}^{(p)}_R &=\hat{y}(H^{(s)}_{R_1^{-}}  e^{i(\vec{k}_{L,i}-\omega t)}+H^{(s)}_{R_1^{+}}  e^{i(\vec{k}_{L,r}-\omega t)}).
\end{split}
\end{equation}

Accordingly, we observe that the total waves at s- or p- polarization can be symbolically described by
\begin{equation}\label{pt1}
\Psi(\vec{r})=\sum_{n=1}^{2}\psi_n u_n^{in}(\vec{r})+\phi_n u_n^{out}(\vec{r})
\end{equation}  
here $\Psi(\vec{r})$ is a wave function that can be $\vec{E}^{(s)}$ or $\vec{H}^{(p)}$, $u_n^{in}(\vec{r})$ and $u_n^{out}(\vec{r})$ represent incoming and outgoing plane waves (modes),  and $\psi_n$ and $\phi_n$ denote the corresponding complex amplitudes.
In the scattering matrix language, we straightforwardly formulate
\begin{equation}
\overline{S}\vec{\psi}=\vec{\phi}
\end{equation}
where $\overline{S}$ is scattering matrix, $\vec{\psi}=[\psi_1, \psi_2]^T$ is incoming wave vector, and $\vec{\phi}=[\phi_1, \phi_2]^T$ is outgoing wave vector.

Following the approach of Ref. \cite{yidong1} and
applying both parity and time reversal operators to Eq. (\ref{pt1}), we obtain
\begin{equation}\label{pt2}
\begin{split}
PT\Psi(\vec{r})&=\sum_{n=1}^{2}PT(\psi_n u_n^{in}(\vec{r}))+PT(\phi_n u_n^{out}(\vec{r}))\\
&=\sum_{m=1}^{2}(PT\psi_n)_mu_m^{out}(\vec{r})+(PT\phi_n)_m u_m^{in}(\vec{r}).
\end{split}
\end{equation}
Parity operator, dented as $P$, preserves incoming and outgoing properties, while it exchanges waves for both sides, i.e, under P operation, $L\rightarrow R$ or $R\rightarrow L$. 
Time reversal operator is anti-linear operator, corresponding to complex conjugation for complex wave amplitudes, while it exchanges incoming and outgoing plane waves (modes) and remains  waves at same sides.

With Eqs.(\ref{pt1}) and (\ref{pt2}),  we derive the scattering relation in a matrix form:
\begin{equation}
\begin{split}
\overline{S}PT\vec{\phi}&=PT\vec{\psi}\\
\overline{S}PT\overline{S}\vec{\psi}&=PT\vec{\psi}\\
\therefore \overline{S}PT \overline{S}&=  PT\\
\Rightarrow PT \overline{S}PT&=  \overline{S}^{-1}\\
\end{split}
\end{equation}
where we already use $(PT)^2=I$ and $I$ is identity matrix.

\section*{Appendix B}
With the 2D parametric space in terms of $Im[M_{unit}^{\sigma}]_{12}$ and $Im[M_{unit}^{\sigma}]_{21}$, we display the PT phase for the polarization-dependent unit cell at different incident angles by $0.675$, $0.75$, and $1.186$ as shown in Figs. 5 (a), (b), and (c), respectively.
We can see that all s-polarizing unit cells at these incident angles are at reflection band.
Moreover, in the cases of $\theta_{in}=0.75$ and $\theta_{in}=1.186$ as Figs. 5 (b) and (c), the p-polarizing unit cells are at EP. Here the black dashed lines correspond to EP.
However, in the case of $\theta_{in}=0.675$ as Fig. 5 (a),  the p-polarizing unit cell  is not at EP, instead at PT broken symmetry phase.
We then calculate the total transmittance and reflectances with respect with N in Fig. 6.
We can observe that the total transmittance is N-dependent and two reflectances are zero at $N=4$, supporting bi-directional transparency. 

 \begin{figure}[ht]
\centering
\includegraphics[width=0.5\textwidth]{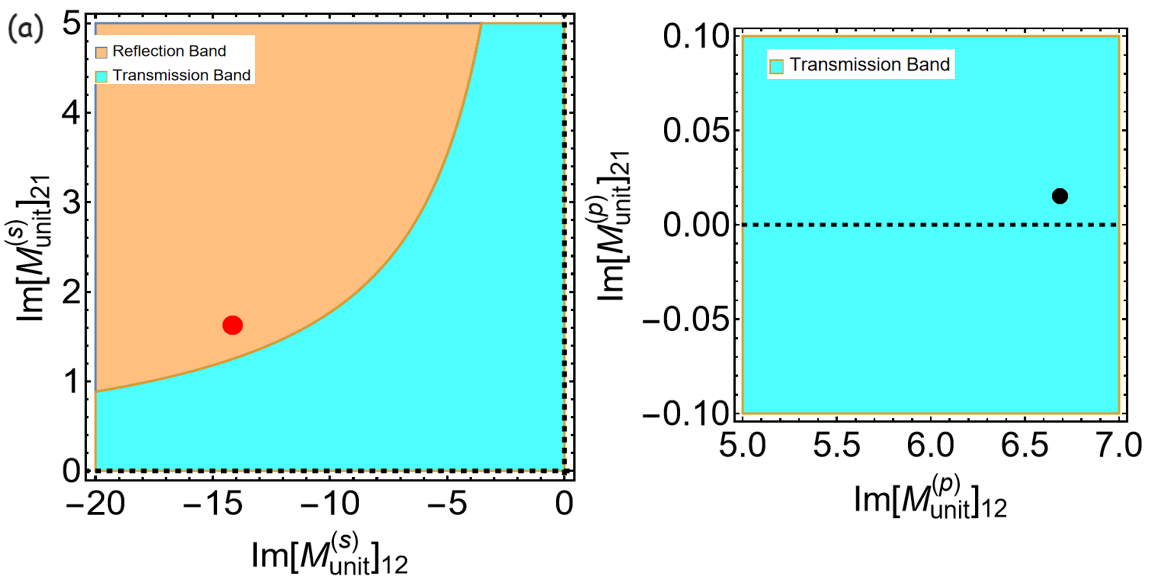}
\includegraphics[width=0.5\textwidth]{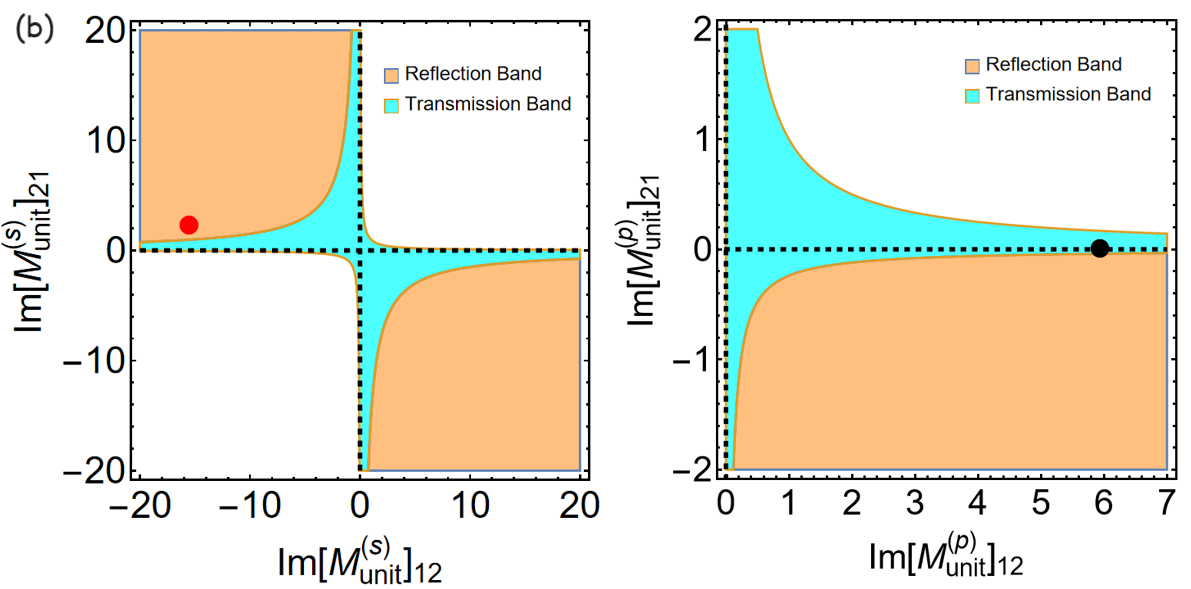}
\includegraphics[width=0.5\textwidth]{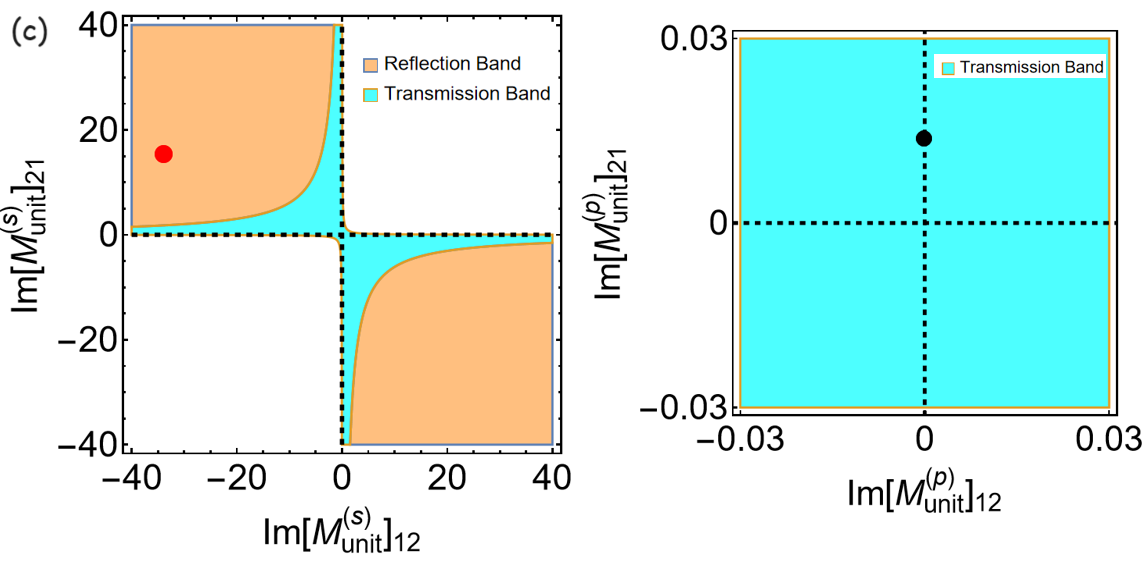}
\caption{ By utilizing the 2D parametric space, we consider the scattering of the polarization-dependent unit cell at incident angles by $0.675$, $0.75$, and $1.186$ in (a), (b), and (c), respectively, marked by red and black dots. Here the black dashed lines mean EP.} 
\end{figure}

 \begin{figure*}[ht]
\centering
\includegraphics[width=1\textwidth]{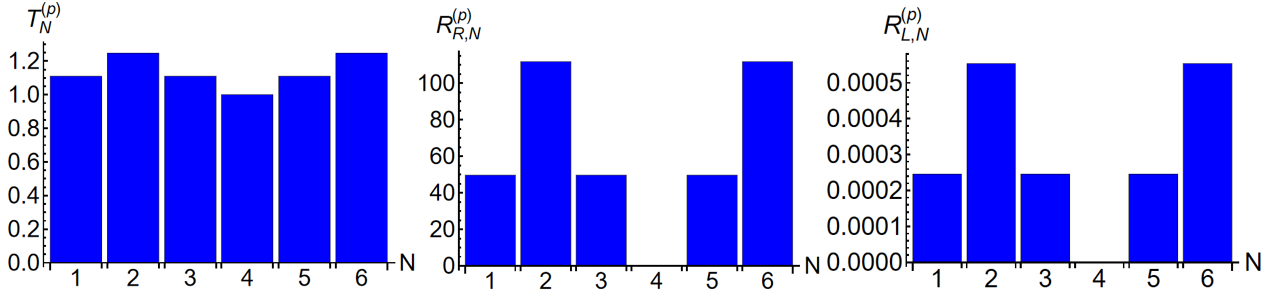}
\caption{ Calculation of the total transmittance and reflectances at $\theta_{in}=0.675$ at p-polarization excitation with respect to N. } 
\end{figure*}

\section*{Appendix C}
The transfer matrix of unit cell is
\begin{equation}
\begin{split}
M_{unit}^{\sigma}=\begin{bmatrix}
\frac{1}{t_1^{\sigma *}} & \frac{r^{\sigma }_{R,1}}{t_1^{\sigma }}\\
-\frac{r^{\sigma}_{L,1}}{t_1^{\sigma }} &  \frac{1}{t_1^{\sigma}}
\end{bmatrix}
\end{split}
\end{equation}
here $t_1^{\sigma }$, $r^{\sigma}_{L,1}$, and $r^{\sigma}_{R,1}$ are complex transmission, left reflection, and right reflection coefficients.
These are polarization dependent coefficients, denoted by a symbol of $\sigma$.
With the help of the Chebyshev identity, the corresponding transfer matrix of a finite periodic system made of N-cells can be expressed by
\begin{widetext}
\begin{equation}
\begin{split}
M_{N}^{\sigma}=\begin{bmatrix}
\frac{1}{t_1^{\sigma *}}\frac{\sin[N\Phi^{\sigma }]}{\sin\Phi^{\sigma }}-\frac{\sin[(N-1)\Phi^{\sigma }]}{\sin\Phi^{\sigma }} & \frac{r^{\sigma }_{R,1}}{t_1^{\sigma }}\frac{\sin[N\Phi^{\sigma }]}{\sin\Phi^{\sigma }}\\
-\frac{r^{\sigma}_{L,1}}{t_1^{\sigma }}\frac{\sin[N\Phi]}{\sin\Phi^{\sigma }} &  \frac{1}{t_1^{\sigma}}\frac{\sin[N\Phi^{\sigma }]}{\sin\Phi^{\sigma }}-\frac{\sin[(N-1)\Phi^{\sigma }]}{\sin\Phi^{\sigma }}
\end{bmatrix}
\end{split}
\end{equation}
  \end{widetext}
here $\Phi^{\sigma }$ is the polarization-dependent Bloch phase in an infinite periodicity.
Accordingly, we can establish a right reflection relation for the unit cell and its N-cell as
\begin{equation}\label{pt6}
\begin{cases}
\begin{split}
\frac{r^{\sigma }_{R,N}}{t_N^{\sigma }}=\frac{r^{\sigma }_{R,1}}{t_1^{\sigma }}\frac{\sin[N\Phi]}{\sin\Phi}\\
\frac{r^{\sigma }_{L,N}}{t_N^{\sigma }}=\frac{r^{\sigma }_{L,1}}{t_1^{\sigma }}\frac{\sin[N\Phi]}{\sin\Phi}.
\end{split}
\end{cases}
\end{equation}
In our case, we require $T_{1}^{(p)}=1$, so $T_{N}^{(p)}=1$. 
Thus we obtain the reflectance relation between finite periodic system and its unit cell, 
\begin{equation}
R_{R,N}^{(p)}=R_{R,1}^{(p)}(\frac{\sin[N\Phi^{(p)}]}{\sin[\Phi^{(p)}]})^2.
\end{equation}

As any finite systems are operated at reflection band,  i.e., $T^{\sigma }_N\rightarrow 0$, with Eq. (\ref{pt6}), we could obtain
\begin{equation}
R^{\sigma}_{R,N}=\frac{R^{\sigma}_{R,1}}{1-T^{\sigma}_1}
\end{equation}
where the term of $\frac{\sin[N\Phi^{\sigma }]}{\sin\Phi^{\sigma }}$ can be replaced by the total transmission of Ref.\cite{finite1},
\begin{equation}
\frac{1}{T^{\sigma}_N}-1=(\frac{1}{T^{\sigma}_1}-1)\frac{\sin^2[N\Phi^{\sigma }]}{\sin^2\Phi^{\sigma }}.
\end{equation}

For any finite periodic PT-symmetry systems, there satisfy the generalized unitarity relation of $\vert T^{\sigma }_N-1 \vert=\sqrt{R_{R,N}^{\sigma }R_{L,N}^{\sigma }}$ \cite{yidong2}.
As a result, at reflection band, there would form a  pair of reciprocal moduli for right- and left-reflection,
\begin{equation}
R_{R,N}^{\sigma }R_{L,N}^{\sigma }=1.
\end{equation}

\end{document}